\documentclass[12pt]{article}
\usepackage{cjp}

\def\lb{\lbrack}
\def\rb{\rbrack}

 \def\Slash#1{
  \begin{picture}(5,6)(0,0)
  \put(-.7,-1.2){\line(5,6)6}
  \end{picture}
  \kern-.8em#1}
 \def\slash#1{
  \begin{picture}(5,6)(0,0)
  \put(-1.5,-1.7){\line(5,6)5}
  \end{picture}
  \kern-.8em#1}

\def\Sn{\Slash \nabla}
\def\sd{\Slash \partial}

\def\Tr{\mbox{Tr}}
\def\tr{\mbox{tr}}
\def\index{\mbox{index}}

\def\be{\begin{eqnarray}}
\def\ee{\end{eqnarray}}

\def\nup{\nabla_{\mu}^+}
\def\num{\nabla_{\mu}^-}
\def\nupm{\nabla_{\mu}^{\pm}}
\def\tA{\widetilde{A}}
\def\tU{\widetilde{U}}
\def\tF{\widetilde{F}}
\def\tC{\widetilde{\cal C}}
\def\wdelta{\widehat{\delta}}
\def\tg5{\widetilde{\Gamma}_5}
\def\hK{\widehat{K}}
\def\dtg5{\widetilde{\Gamma}_5^{K\hK}}
\def\dg5{\Gamma_5^{K\hK}}

\def\g5{\gamma_5}
\def\hg5{\hat{\gamma_5}}

\def\C{{\cal C}}

\begin{document}

\title{Dirac operator index and topology of lattice gauge fields}
\author{David H. Adams}
\address{Dept. of Pure Mathematics, University of Adelaide \\
Adelaide, S.A. 5005, Australia. \\ Email: dadams@maths.adelaide.edu.au}
\date{\today}
\maketitle

\begin{abstract}
The fermionic topological charge of lattice gauge fields, given in terms
of a spectral flow of the Hermitian Wilson--Dirac operator, or equivalently,
as the index of Neuberger's lattice Dirac operator, is shown to have 
analogous properties to L\"uscher's geometrical lattice topological charge.
The main new result is that it reduces to the continuum topological charge
in the classical continuum limit. (This is sketched here; the full proof 
will be given in a sequel to this paper.)
A potential application of the ideas behind fermionic lattice topological 
charge to deriving a combinatorial construction of the signature invariant of 
a 4-manifold is also discussed.

\end{abstract}


\section{Background}

In the continuum, a smooth SU(2) gauge field $A(x)=A_{\mu}^a(x)T^adx^{\mu}$
on the Euclidean hyper-torus $T^4$ is from a mathematical point of view a
connection 1-form on a principal SU(2) bundle $P$ over $T^4$.
The bundle is characterised (up to topological equivalence) by an integer
$Q_P$ and this number can be recovered from the gauge field:
\be
Q_P=Q(A)=\frac{1}{8\pi^2}\int_{T^4}\tr(F\wedge{}F)=
\frac{1}{32\pi^2}\int{}d^4x\,\epsilon_{\mu\nu\rho\sigma}
\tr(F_{\mu\nu}(x)F_{\rho\sigma}(x))
\label{1}
\ee
where $F=dA+\frac{1}{2}[A,A]$ is the curvature of $A$. Thus the topological 
structure of $P$ is encoded in the gauge field $A$. 
It is also encoded in the space of zero-modes of
the Dirac operator $\sd_A=\gamma^{\mu}(\partial_{\mu}+A_{\mu})$:
the Atiyah--Singer index theorem \cite{AS} gives
\be
\index(\sd_A)\;\equiv\;\Tr(\gamma_5\Big|_{\ker\sd_A})=Q(A)\,.
\label{2}
\ee
The space of all SU(2) gauge fields on $T^4$ is a disjoint union of components
(topological sectors) labelled by $Q\in{\bf Z}$.

Now put a lattice (i.e. hyper-cubic cell decomposition) on $T^4$ with
lattice spacing $a$. In \cite{L} M. L\"uscher showed that a lattice gauge
field $U_{\mu}(x)\,$, defined on the links $[x,x+ae_{\mu}]$ 
($e_{\mu}$=unit vector on the positive $\mu$-direction) and taking values
in SU(2), also has encoded in it the structure of a principal SU(2) bundle
$P$ over $T^4$: transition functions for $P$ on the overlaps of a collection
of regions covering $T^4$ were explicitly constructed from $U$.
An integer topological charge for $U$ is then given by $Q^{geo}(U)=Q_P$.
This is often referred to as the geometrical lattice topological charge
(to distinguish it from the fermionic topological charge discussed below).
The construction of $P$ from $U$ is ambiguous for certain ``exceptional''
lattice gauge fields; roughly speaking, these are the fields for which it is
not possible to canonically write $U(p)=\exp(\tau(p))$ for each plaquette
$p\,$, where $U(p)$ is the product of the link variables around $p$.
(Example: in the analogous case of U(1) gauge fields on the 2-torus
\cite{PhillipsU(1)}, $U(p)\in\mbox{U(1)}$ can be canonically written as 
$e^{\tau(p)}\,$, $i\tau(p)\in(-\pi,\pi)$ except for the exceptional fields
which have $U(p)=-1$ for some $p$.)
The exceptional fields form a lower-dimensional manifold in the space of 
lattice gauge fields. They can be excluded by restricting to the lattice gauge
fields whose plaquette variables 
are sufficiently close to the identity $1$ in SU(2); a sufficient
condition is \cite{L}\footnote{For 
$U\in\mbox{SU(2)}$ we have $\tr(1-U)\,\le\,2||1-U||$.}
\be
\tr(1-U(p))\;<\;0.03\qquad\mbox{or}\qquad{}||1-U(p)||\;<\;0.015
\label{3}
\ee
where we are using the matrix norm defined by 
$||M||^2=\frac{1}{d}\sum_{I}|M_I|^2$ for $d\times{}d$ matrix $M$ with the sum
running over the column vectors $M_I$ of $M$ (the normalisation factor
$\frac{1}{d}$ is so that unitary matrices have norm 1).
For the non-exceptional fields the bundle $P$ is uniquely determined so
$Q^{geo}(U)=Q_P$ is well-defined. Furthermore, the topological structure of $P$
is unchanged under continuous deformation of $U$ provided that no exceptional
fields are encountered \cite{L}. Thus after excising the exceptional fields
the space of lattice gauge fields acquires a non-trivial topological structure,
decomposing into disjoint topological sectors which are again labelled by
$Q\in{\bf Z}$. In the continuum $Q$ can take any value in ${\bf Z}\,$, but
in the lattice setting $Q$ can only take values in a finite subset of 
${\bf Z}$ depending on how fine the lattice is \cite{PhillipsU(1),PS}.
When $U$ is the lattice transcript of a smooth continuum field $A$ 
(and $U$ is non-exceptional) the bundle $P_U$ specified by $U$ need not
coincide with the bundle $P_A$ specified by $A$ in general. However, 
L\"uscher showed that $P_U$ and $P_A$ do coincide when the lattice is
sufficiently fine: he showed that $Q^{geo}(U)\to{}Q(A)$ in the classical 
continuum limit $a\to0$ \cite{L}.
The topology of lattice gauge fields was further elucidated by A. Phillips 
and D. Stone in \cite{PhillipsU(1),PS} where $Q^{geo}(U)$ was also generalised
to the case of SU(N) gauge fields.

In light of the Index Theorem (\ref{2}) in the continuum, it is natural
to ask if there is a lattice Dirac operator $D$ such that
\be
\index(D_U)\stackrel{?}{=}Q^{geo}(U)
\label{4}
\ee
Such a lattice version of the Index Theorem would certainly be mathematically
interesting. It would also be of physical relevance: 't Hooft's
solution of the U(1) problem \cite{U(1)problem} uses in a crucial way
the connection between the topological structure of $A$ and zero-modes of
$\sd_A$ implied by the Index Theorem (\ref{2}). Also, an alternative
``fermionic'' description of $Q^{geo}(U)$ as the index of a lattice Dirac operator
could be practically useful for numerical work, seeing as the expression for
$Q^{geo}(U)$ is quite complicated and time-consuming to implement
numerically.
Traditional lattice Dirac operators are a bit problematic in this context.
To avoid doubler zero-modes, operators such as the 
Wilson--Dirac operator \cite{Wilson}
include a chiral symmetry-breaking term; this results in the nullspace
$\ker{}D$ not being invariant under $\g5$ so the zero-modes do not have 
definite chirality and $\index\,D$ is not defined.
Nevertheless, a lattice version of the index can still be defined from
the eigenvectors of the Wilson--Dirac operator $D_U^{Wilson}$ with low-lying
real eigenvalues \cite{Smit}. Essentially the same lattice index is obtained
\cite{Itoh} as minus the spectral flow of the Hermitian Wilson--Dirac operator
\be
H_U(m)=\g5(D_U^{Wilson}-{\textstyle \frac{m}{a}})
\label{5}
\ee
coming from the crossings of the origin by eigenvalues $\lambda_k(m)$
at low-lying values of $m$.\footnote{The notation in \cite{Itoh}
is different; instead of $m$ they have a different parameter $K$.} 
\footnote{In the continuum, taking the $\gamma$-matrices to be hermitian
so $\sd_A$ is anti-hermitian, crossings of the origin by eigenvalues of
the hermitian operator $H_A(m)=\g5(\sd_A-m)$ only happen at $m=0$ and the 
spectral flow from these is $-\,\index(\sd_A)$.}
(Note that if $\lambda_k(m_0)=0$ then the corresponding eigenvector 
$\psi_k(m_0)$ is an eigenvector for $D_U^{Wilson}$ with eigenvalue 
$\frac{m_0}{a}$.) This ``fermionic'' definition of the topological charge 
of a lattice gauge field, which we denote by $Q^f(U)\,$, 
can be made more precise by defining $Q^f(U)$ to be minus the spectral
flow of $H_U(m)$ as $m$ varies from 0 to 1; this is justified 
by the fact \cite{Itoh}
that for ``sufficiently smooth'' lattice gauge fields the real eigenvalues
of $D_U^{Wilson}$ are positive and are localised around $\frac{m}{a}\,$, 
$m=0,2,4,6,8$ (an analytic explanation of this was 
recently sketched in \cite{DA(herm)}).
Note that $Q^f(U)$ has properties analogous to those of 
$Q^{geo}(U)$ discussed above: It is defined for all $U$ except those for which
$H_U(1)$ has a zero-mode; these are exceptional in the sense that they form
a lower-dimensional manifold in the space of lattice gauge fields.
It is clear from the definition in terms of
spectral flow that $Q^f(U)$ is constant under continuous variations
of $U$ provided that no exceptional fields are encountered. 
Thus $Q^f(U)$ determines a topological structure in the space of lattice
gauge fields in the same way as $Q^{geo}(U)$ . The exceptional fields for 
$Q^f(U)$ can be excluded by imposing a condition on the plaquette 
variables \cite{local,Neu(bound)}:
\be
||1-U(p)||\;<\;0.03
\label{8}
\ee
Note the similarity with (\ref{3}). (Neither (\ref{3}) nor (\ref{8})
are optimal.)

$Q^f(U)$ has subsequently appeared in 
other guises, as the overlap topological charge in \cite{overlap} and more 
recently as the index of a new lattice Dirac operator: Neuberger's 
Overlap-Dirac operator \cite{Neu1}, given by
\be
D_U=\frac{1}{a}\biggl(\,1+\g5\frac{H_U}{\sqrt{H_U^2}}\biggr)
\label{6}
\ee
where $H_U\equiv{}H_U(1)$. This operator satisfies \cite{Neu2} the 
Ginsparg--Wilson relation \cite{GW} $D\g5+\g5{}D=aD\g5{}D\,$, which
implies that $\ker{}D$ is invariant under $\g5$ 
(since $D\psi=0$ $\Rightarrow$
$D(\g5\psi)=(aD\g5{}D-\g5{}D)\psi=0$) so the zero-modes of $D$ have definite
chirality and $\index\,D=\Tr(\g5|_{\ker{}D})$ is well-defined, as was 
first noted in \cite{laliena}. A formula for the index gives
\cite{laliena,Luscher(PLB),Fuji(NPB)}
\be
\index(D_U)=-\frac{a}{2}\Tr(\g5{}D_U)=-\frac{1}{2}\Tr\Bigl(\,
\frac{H_U}{\sqrt{H_U^2}}\Bigr)=Q^f(U)
\label{7}
\ee
where the last equality follows from the facts that $\Tr\Bigl(\,
\frac{H_U}{\sqrt{H_U^2}}\Bigr)$ is the spectral asymmetry of $H_U(m)$ 
at $m=1$ and
$H_U(m)$ has symmetric spectrum for $m<0$ \cite{overlap,Edwards}.

The fermionic lattice topological charge, in its various guises, has been
an ongoing subject of study since the original works 
\cite{Smit,Itoh,Seiler(Top)} in the mid 1980's; for a selection of recent 
works see, e.g., \cite{Vranas,Edwards,Gattringer,Hernandez,Chiu(PRD)(PRD)}. 
The emphasis has tended to be on numerical investigations though,
and there are a number of interesting questions which have been partially
answered by numerically studies but which currently lack a complete analytical
resolution. Perhaps the most fundamental of these is the question of whether
$Q^f$ and $Q^{geo}$ are equal (at least under suitable conditions on the
lattice gauge field, such as (\ref{3}),(\ref{8})).\footnote{This question
has received attention in numerical studies, e.g. in \cite{Itoh} where 
$Q^f(U)$ was compared to a simpler version of $Q^{geo}(U)$ due to 
P. Woit \cite{Woit}.}
This is equivalent to the question of whether the Lattice Index Theorem 
(\ref{4}) holds with $D_U$ being the Overlap-Dirac operator (\ref{6}).
Another basic question is whether $Q^f(U)$ reduces to the continuum 
topological charge $Q(A)$ (= $\index(\sd_A)\,$) in the classical continuum 
limit. (This is a necessary condition for equality between $Q^f$ and $Q^{geo}$ 
since, as discussed above, $Q^{geo}(U)\to{}Q(A)$ in this limit \cite{L}.)
This has sometimes been taken for granted in the literature; e.g. in 
\cite{Itoh} it is claimed (without proof) that the eigenvectors of 
$D_U^{Wilson}$ with low-lying 
real eigenvalues will ``reduce to'' the zero-modes
of $\sd_A$ in the classical continuum limit. To give a precise mathematical
formulation and proof of this statement is not so easy though. From a 
mathematical point of view, showing $Q^f(U)\to{}Q(A)$ in the
classical continuum limit is a long-standing open problem, and the main 
purpose of this article is to announce and sketch a proof of this:

\vspace{1ex}
\noindent {\em Theorem.} For SU(N) gauge fields on the Euclidean 4-torus,
$Q^f(U)\to{}Q(A)$ in the classical continuum limit. Equivalently,
$\index(D_U)\to\index(\sd_A)$ in this limit, where $D_U$ is the 
Overlap-Dirac operator (\ref{6}) above.
\vspace{1ex}

This is the fermionic analogue of L\"uscher's result $Q^{geo}(U)\to{}Q(A)$.
The deeper question of whether $Q^f$ and $Q^{geo}$ are  
equal (at least when $U(p)$ satisfies a bound of the form
(\ref{3}),(\ref{8})) is a challenging 
mathematical problem which we will not attempt here.

\section{Classical continuum limit of the fermionic lattice topological
charge}

The proof of the theorem above, which we sketch in the following, grew out of 
suggestions by Martin L\"uscher \cite{Luscher(private)}. The full details
will be given in a forthcoming paper \cite{prep}. (An alternative
argument for a restricted class of topologically non-trivial fields in
a slightly different setting\footnote{which can be reduced to the present
setting \cite{Luscher(private)}} was previously given in 
\cite[v4]{DA(hep-lat)}.)
The general idea is to start with the last equality in (\ref{7}),
\be
Q^f(U)=-\frac{1}{2}\Tr\Bigl(\,\frac{H_U}{\sqrt{H_U^2}}\Bigr)\,,
\label{8.5}
\ee
carry out a certain power series expansion of the inverse square root,
and then evaluate the trace in an explicit basis for the spinor fields.
But we do this in a slightly indirect way: First, we use the locality
result of \cite{local} for the Overlap-Dirac operator to
derive a relation
$Q^f(U)=Q^f(U)^{(2p+1)}+O(e^{-c/a})$ (with $c>0$) where 
$Q^f(U)^{(2p+1)}$ is the fermionic topological charge in a setting in which 
an infinite volume limit $p\to\infty$ can be taken. 
Then $\lim_{a\to0}\,Q^f(U)=\lim_{a\to0}\,\lim_{p\to\infty}\,Q^f(U)^{(2p+1)}$
and the latter quantity is easier to evaluate because the sum resulting 
from the trace in (\ref{8.5}) becomes a tractable integral in the 
$p\to\infty$ limit.

\subsection{Preliminaries}
The 4-torus $T^4$ is taken to be 
$[-\frac{1}{2}L,\frac{1}{2}L]/_{\sim}\times\cdots\times
[-\frac{1}{2}L,\frac{1}{2}L]/_{\sim}$ (where $\sim$ means identify endpoints). 
Then a continuum SU(2)
gauge field $A_{\mu}(x)$ on $T^4$ can be viewed as a gauge field on
${\bf R}^4$ satisfying 
\be
A_{\mu}(x+Le_{\nu})=\Omega(x,\nu)A_{\mu}(x)\Omega(x,\nu)^{-1}
+\Omega(x,\nu)\partial_{\mu}\Omega(x,\nu)^{-1}
\label{9}
\ee
where $\Omega(x,\nu)\,$, $\nu=1,2,3,4\,$, are the SU(2)-valued monodromy
fields which specify the principal SU(2) bundle over $T^4$.\footnote{
These also satisfy a cocycle condition which ensures that 
$A_{\mu}(x+Le_{\nu}+Le_{\rho})$ is unambiguous. It is always possible to
make a gauge transformation so that $\Omega(x,\nu)=1$ for $\nu=1,2,3$
and $\Omega(x,4)$ is periodic in $x_1,x_2,x_3$. Then for fixed $x_4$
$\Omega(x,4)$ determines a map $T^3\to\mbox{SU(2)}$. The degree of this map
(which is independent of $x_4$ since $\Omega(x,4)$ depends smoothly on $x_4$)
is precisely the integer $Q_P$ specifying the SU(2) bundle $P$ over $T^4\,$,
and is therefore also the topological charge $Q(A)=Q_P$ of $A$.}
Now put a hyper-cubic lattice on $T^4$ with $2N$ sites along each edge, so
the lattice spacing is $a=L/2N$. This extends to a hyper-cubic lattice
on ${\bf R}^4$ with sites $a{\bf Z}^4$. The lattice transcript of $A\,$,
\be
U_{\mu}(x)=T\exp\Bigl(\,\int_0^1\,aA_{\mu}(x+tae_{\mu})\,dt\Bigr)
\label{10}
\ee
(T=t-ordering and for simplicity the coupling constant has been set to
unity) satisfies
\be
U_{\mu}(x+Le_{\nu})=\Omega(x,\nu)U_{\mu}(x)\Omega(x+ae_{\mu},\nu)^{-1}\,.
\label{11}
\ee
The finite-dimensional complex vector-space $\C$ of lattice spinor fields
on $T^4$ consists of the spinor fields $\psi(x)\,$, $x\in{}a{\bf Z}^4\,$, 
satisfying
\be
\psi(x+Le_{\nu})=\Omega(x,\nu)\psi(x)\,.
\label{12}
\ee
The covariant finite difference operators $\frac{1}{a}\nupm\,$, given by
\be
\nup\psi(x)&=&U_{\mu}(x)\psi(x+ae_{\mu})-\psi(x) \label{13} \\
\num\psi(x)&=&\psi(x)-U_{\mu}(x-ae_{\mu})^{-1}\psi(x-ae_{\mu})\,, 
\label{14}
\ee
preserve (\ref{12}) and are therefore well-defined on $\C\,$,
as is the Wilson--Dirac operator, given by\footnote{For simplicity
we have taken the Wilson parameter to be $r=1$.}
\be
D_U^{Wilson}={\textstyle \frac{1}{a}}
\Sn_U+{\textstyle \frac{1}{2}}a\Bigl(
{\textstyle \frac{1}{a^2}}\Delta_U\Bigr)
\label{14.5}
\ee
where $\frac{1}{a}\Sn=\frac{1}{a}\sum_{\mu}\gamma^{\mu}\nabla_{\mu}=
\frac{1}{a}\sum_{\mu}\gamma^{\mu}\frac{1}{2}(\nup+\num)$ is the naive
lattice Dirac operator (the $\gamma^{\mu}$'s are taken to be hermitian
so $\Sn$ is anti-hermitian), $\frac{1}{a^2}\Delta=\frac{1}{a^2}\sum_{\mu}
\num-\nup=\frac{1}{a^2}\sum_{\mu}(\nup)^*\nup
=\frac{1}{a^2}\sum_{\mu}(\num)^*\num$ is the lattice Laplace operator
(hermitian, positive). Likewise $H_U=\g5(D_U^{Wilson}-\frac{1}{a})$ 
is well-defined on $\C\,$, and so is 
the Overlap-Dirac operator $D_U=\frac{1}{a}(1+\g5H_U|H_U|^{-1})$ provided
that $H_U$ has no zero-modes.
$H_U$ is guaranteed not to have zero-modes when $a$ is sufficiently 
small. Indeed, 
$1-U(p_{x,\mu,\nu})=-a^2F_{\mu\nu}(x)+O(a^3)$ vanishes uniformly 
for $a\to0\,$, implying that (\ref{8}) is satisfied for sufficiently small 
$a\,$, which in turn implies that \cite{local,Neu(bound)}
\be
H_U^2\;\ge\;b\;>\;0
\label{14.7}
\ee
(This holds, e.g., with $b=0.1$ but 
we will not need the explicit value in the following.)
We henceforth assume $a$ to be small enough that (\ref{8}) --
and thereby (\ref{14.7}) -- holds.

\subsection{Passing to a setting with an infinite volume limit}

A general linear operator ${\cal O}$ on lattice spinor fields corresponds
to a kernel function ${\cal O}(x,y)$ via ${\cal O}\psi(x)=
a^4\sum_y{\cal O}(x,y)\psi(x)$. By (\ref{7}) we have
\be
Q^f(U)=a^4\sum_{x\in\Gamma}q_U(x)
\label{15}
\ee
where
\be
q_U(x)=-\frac{a}{2}\tr(\g5{}D_U(x,x))
\label{16}
\ee
and the summation is over
\be
\Gamma=\{x=an\,|\,n_{\mu}=-N,-N+1,\dots,N-1\}\,.
\label{17}
\ee
Now, for arbitrary whole number $p\,$, set
\be
\Omega^{(p)}(x,\nu)=\Omega(x+pLe_{\nu},\nu)\Omega(x+(p-1)Le_{\nu}, \nu)
\cdots\Omega(x+Le_{\nu},\nu)\Omega(x,\nu)
\label{18}
\ee
and define $\C_p$ to be the space of lattice spinor fields $\psi(x)\,$,
$x\in{}a{\bf Z}^4\,$, satisfying
\be
\psi(x+(p+1)Le_{\nu})=\Omega^{(p)}(x,\nu)\psi(x)\,.
\label{19}
\ee
Note that (\ref{12}) implies (\ref{19}), so $\C$ is contained in
$\C_p$ for all $p$. Note also that the covariant finite difference operators
(\ref{13})--(\ref{14}) preserve (\ref{19}) and are therefore well-defined 
operators on $\C_p\,$; it follows that the Overlap-Dirac operator is 
well-defined as an operator on $\C_p$. We denote the Overlap-Dirac operator
on $\C_p$ by $D_U^{(p)}$ in the following, with $D_U$ denoting the 
operator on $\C$. The fact that 
$D_U$ is the restriction of $D_U^{(p)}$ to $\C\subseteq\C_p$ for
all $p$ implies
\be
D_U\psi(x)=D_U^{(2p+1)}\psi(x)=a^4\sum_{y\in\Gamma_{2p+1}}
D_U^{(2p+1)}(x,y)\psi(y)\qquad\mbox{for}\ \ \psi\in\C
\label{20}
\ee
where
\be
\Gamma_{2p+1}&=&\{x=an\,|\,n_{\mu}=-(2p+1)N,-(2p+1)N+1,\dots,(2p+1)N-1\} 
\nonumber \\
&=&\{x+Lm\,|\,x\in\Gamma\ ,\ m_{\mu}=-p,-p+1,\dots,p\}\,.
\label{21}
\ee
From this it is straightforward to derive \cite{prep}, using (\ref{12}), that
for $p\ge1$ the norm of
\be
R_U^{(2p+1)}(x,y):=D_U(x,y)-D_U^{(2p+1)}(x,y)
\label{22}
\ee
has a bound
\be
||R_U^{(2p+1)}(x,y)||\;\le\;\sum_{|m_{\mu}|\in\{1,2,\dots,p\}}
||D_U^{(2p+1)}(x,y+Lm)||\,.
\label{23}
\ee
The locality result in \cite{local} now gives
\be
||aD_U^{(2p+1)}(x,x+Lm)||\;\le\;\frac{c_1}{a^4}\,\exp\Bigl(-c_2\frac{L}{a}
\sum_{\mu}|m_{\mu}|\Bigr)
\label{24}
\ee
where $c_1$ and $c_2>0$ are constants independent of $a$, $p$ and $U$.
It follows that 
\be
||aR_U^{(2p+1)}(x,x)||&\le&2^4\sum_{m_{\mu}\in\{1,2,\dots,p\}}
\frac{c_1}{a^4}\,\exp\Bigl(-c_2\frac{L}{a}\sum_{\mu}m_{\mu}\Bigr)
\nonumber \\
&\le&2^4\frac{c_1}{a^4}\prod_{\mu}\int_{1/2}^{p+1/2}dt_{\mu}\,
\exp\Bigl(-c_2\frac{L}{a}t_{\mu}\Bigr)
\nonumber \\
&\le&c_1\Bigl(\frac{2}{c_2L}\Bigr)^4\,\exp\Bigl(-\frac{c_2L}{2a}\Bigr)\,.
\label{26}
\ee
Set 
\be
q_U^{(2p+1)}(x)=-\frac{a}{2}\tr(\g5{}D_U^{(2p+1)}(x,x))
\label{28}
\ee
then substituting (\ref{22}) in (\ref{16}) and taking account of (\ref{26})
we see that the norm of
\be
r_U^{(2p+1)}(x):=q_U(x)-q_U^{(2p+1)}(x)
\label{27}
\ee
has a bound of the form
\be
|r_U^{(2p+1)}(x)|\;\le\;O(e^{-c/a})\quad\quad\ c>0
\label{29}
\ee
where $O(e^{-c/a})$ is independent of $U$ and $p$. It follows that
\be
\lim_{a\to0}\;q_U(x)&=&\lim_{a\to0}\,\lim_{p\to\infty}\;q_U^{(2p+1)}(x)
\qquad\quad(x\in\Gamma)
\label{30a} \\
\lim_{a\to0}\;Q^f(U)&=&
\lim_{a\to0}\,\lim_{p\to\infty}\;a^4\sum_{x\in\Gamma}q_U^{(2p+1)}(x)
\label{30}
\ee
The infinite volume limit $p\to\infty$ in these expressions will facilitate
their evaluation, as we will see in the following subsection.

\subsection{Evaluation of the classical continuum limit}

We now exploit the locality of the Overlap-Dirac operator in the gauge field
\cite{local} to replace $A_{\mu}(x)$, or rather, its lattice transcript 
$U_{\mu}(x)$, in (\ref{30a})--(\ref{30}) by the lattice transcript 
$\tU_{\mu}(x)$ of a gauge field $\tA_{\mu}(x)$ 
which coincides with $A_{\mu}(x)$ in a neighbourhood of 
$[-\frac{1}{2}L,\frac{1}{2}L]^4$ 
but which vanishes outside a bounded region in ${\bf R}^4$.
Specifically, choose a 
smooth function $\lambda(x)$ on ${\bf R}^4$ which equals 1
in a neighbourhood of $[-\frac{1}{2}L,\frac{1}{2}L]^4$ and vanishes outside a 
region contained in $[-\frac{3}{2}L,\frac{3}{2}L]^4\,$, and 
set $\tA_{\mu}(x)=\lambda(x)A_{\mu}(x)$. 
For each $p\ge1$ we take $\tU$ to be the lattice transcript of $\tA$ in 
$[-\frac{2p+1}{2}L,\frac{2p+1}{2}L]^4$ and extend 
$\tU$ to the rest of the lattice on ${\bf R}^4$ 
by requiring that $\tU_{\mu}(x)$ be periodic in all directions with
period $(2p+1)L$.
Then the Overlap-Dirac operator with lattice gauge field $\tU$ is 
a well-defined operator $D_{\tU}^{(2p+1)}$ on the space $\tC_{2p+1}$ of 
lattice spinor fields satisfying the periodicity condition 
$\psi(x+(2p+1)Le_{\nu})=\psi(x)\,$, $\nu=1,2,3,4$.
A version of the locality result of \cite{local} leads to \cite{prep}
\be
||aD_U^{(2p+1)}(x,x)-aD_{\tU}^{(2p+1)}(x,x)||\;\le\;
\frac{1}{a^4}\,O(e^{-\tilde{c}/a})\qquad\quad\ x\in\Gamma\quad\ p\ge1
\label{31}
\ee
where $\tilde{c}>0$ is a constant and $O(e^{-\tilde{c}/a})$ is independent
of $x\,$, $p\,$, $U\,$, $\tU$. It follows that (\ref{30a})--(\ref{30})
are unchanged if $q_U^{(2p+1)}(x)$ is replaced by 
\be
q_{\tU}^{(2p+1)}(x)=-\frac{1}{a}\tr(\g5{}D_{\tU}^{(2p+1)}(x,x))\,.
\label{33}
\ee
There are several reasons for making the replacement $A\to\tA\,$, 
$U\to\tU$. (i) The rigorous justification in \cite{prep}
of the steps that follow uses the fact that $\tA_{\mu}(x)$ vanishes outside a 
bounded region. (This is not the case in general for 
$A_{\mu}(x)$ which can diverge in the $|x|\to\infty$ limit.)
(ii) It leads to $\C_p$ being replaced by a space $\tC_p$ of periodic 
spinor fields, thereby allowing the trace in (\ref{35}) below to be evaluated
in a ``plane wave'' basis.

Substituting the expression for the Overlap-Dirac operator in (\ref{33})
we find (cf. (\ref{7}))
\be
q_{\tU}^{(2p+1)}(x)&=&-\frac{1}{2}\tr\Bigl(\,\frac{H_{\tU}}{\sqrt{H_{\tU}^2}}
(x,x)\Bigr) \nonumber \\
&=&-\frac{1}{2}\tr\Bigl(\,\g5\Big\{X_{\tU}\frac{1}
{\sqrt{X_{\tU}^*X_{\tU}}}\Big\}(x,x)\Bigr) \nonumber \\
&=&\frac{-1}{2a^4}\Tr\Bigl(\,\g5X_{\tU}\frac{1}{\sqrt{X_{\tU}^*X_{\tU}}}
\,\wdelta_x\Bigr)
\label{35}
\ee
where
\be
X_{\tU}&=&a(D_{\tU}^{Wilson}-{\textstyle \frac{1}{a}}) \nonumber \\
&=&\Sn_{\tU}+{\textstyle \frac{1}{2}}(\Delta_{\tU}-2)
\label{36}
\ee
and the operator
$\wdelta_x$ is defined by $(\wdelta_x\psi)(y)=\psi(x)\delta_{xy}$.
The strategy now is to carry out a power series expansion of the 
inverse square root in (\ref{35}). A calculation gives
\be
X_{\tU}^*X_{\tU}=L_{\tU}+V_{\tU}
\label{38}
\ee
where
\be
L&=&-\nabla^2+{\textstyle \frac{1}{2}}(\Delta-2)^2 \label{39} \\
V&=&-{\textstyle \frac{1}{4}}\lb\gamma^{\mu}\,,\gamma^{\nu}\rb
\lb\nabla_{\mu}\,,\nabla_{\nu}\rb-
{\textstyle \frac{1}{2}}\lb\Sn\,,\Delta\rb\,.
\label{40}
\ee
$V$ is a linear combination of commutators of the $\nupm$'s. As pointed out 
in \cite{local}, the norms of these commutators are bounded by 
$\mbox{max}_p||1-\tU(p)||$. In the present case, this together
with the expansion 
\be
\tU(p_{x,\mu,\nu})=1+a^2\tF_{\mu\nu}(x)+O(a^3)
\label{41}
\ee
show that
\be 
||V_{\tU}||\;\sim\;O(a^2)
\label{42}
\ee
As in (\ref{14.7}) we have a bound $X_{\tU}^*X_{\tU}\ge{}b>0$ when $a$ is
sufficiently small; furthermore, due to (\ref{42}) we can assume that
$||V||\le{}b/2$. Then from (\ref{38}) we get a  
lower bound $L_{\tU}\ge{}b/2>0$ on the positive hermitian operator $L_{\tU}$.
It follows that $L$ is invertible and $||L^{-1}V||<1$ when $a$ is sufficiently
small. The inverse square root in (\ref{35}) can then be expanded as
\be
\frac{1}{\sqrt{X^*X}}&=&\int_{-\infty}^{\infty}\frac{d\sigma}{\pi}\,
\frac{1}{X^*X+\sigma^2} \nonumber \\
&=&\int_{-\infty}^{\infty}\frac{d\sigma}{\pi}\,
\Bigl(\,\frac{1}{L+\sigma^2}\Bigr)\Bigl(\,\frac{1}{1+(L+\sigma^2)^{-1}V}
\Bigr) \nonumber \\
&=&\int_{-\infty}^{\infty}\frac{d\sigma}{\pi}\,\frac{1}{L+\sigma^2}\,
\sum_{k=0}^{\infty}(-1)^k((L+\sigma^2)^{-1}V)^k\,.
\label{43}
\ee
Note that the $\gamma$-matrices in (\ref{43}) are all contained in $V$.
Since the trace of $\g5$ times a product of less than 4 $\gamma$-matrices
vanishes, the terms with $k=0$ and $k=1$ in (\ref{43}) give vanishing
contributions to (\ref{35}). On the other hand, by (\ref{42}) the terms 
with $k\ge3$ are $O(V^3)\sim{}O(a^6)\,$, hence the contribution of these
in (\ref{35}) is $\sim{}O(a^2)$. (This is rigorously established in 
\cite{prep} using the presence of $\wdelta_x$ in (\ref{35}) together with
the fact that $\tA_{\mu}(x)$ has compact support.) Thus the only relevant
contribution to (\ref{35}) from the expansion (\ref{43}) comes from the
$k=2$ term:
\be
\int_{-\infty}^{\infty}\frac{d\sigma}{\pi}\,\frac{1}{(L+\sigma^2)^2}
V\frac{1}{L+\sigma^2}V
\label{44}
\ee
By making a power series expansion of $1/(L+\sigma^2)$ and using
$\lb{}L,V\rb\sim{}O(a)$ we find in \cite{prep} that, modulo an $O(a)$ term,
the contribution of (\ref{44}) in (\ref{35}) is the same as that of
\be
V^2\int_{-\infty}^{\infty}\frac{d\sigma}{\pi}\,\frac{1}{(L+\sigma^2)^3}
=V^2L^{-5/2}\int_{-\infty}^{\infty}\frac{d\sigma}{\pi}\,
\frac{1}{(1+\sigma^2)^3}=\frac{3}{8}V^2L^{-5/2}
\label{45}
\ee
We hereby see that (\ref{35}) reduces to
\be
q_{\tU}^{(2p+1)}(x)=\frac{-3}{16a^4}\Tr(\g5{}X_{\tU}V_{\tU}^2L_{\tU}^{-5/2}
\wdelta_x)+O(a)
\label{46}
\ee
Now, from (\ref{10}) with $A\to\tA$ we see that the $a$-dependence of $\tU$
enters through a product $a\tA$. It follows that 
\be
X_{\tU}=X_1+{\cal O}(a)\qquad\quad,\qquad\ L_{\tU}=L_1+{\cal O}(a)
\label{46.5}
\ee
Since $V\sim{\cal O}(a^2)$ we can replace 
$X_{\tU}\to{}X_1$ and $L_{\tU}\to{}L_1$
in (\ref{46}) at the expense of another $O(a)$ term. With the expressions
(\ref{36}) and (\ref{39}) for $X$ and $L$ we find
\be
q_{\tU}^{(2p+1)}(x)=\frac{-3}{16a^4}
\Tr\Bigl(\,\g5\wdelta_xV_{\tU}^2(\Sn_1+{\textstyle \frac{1}{2}}
(\Delta_1-2))(-\nabla_1^2+
{\textstyle \frac{1}{2}}(\Delta_1-2))^{-5/2}\Bigr)+O(a)
\label{47}
\ee
The rigorous derivation of this in \cite{prep} again uses the fact that 
$\tA_{\mu}(x)$ has compact support.
The trace in (\ref{47}) can now be evaluated using the ``plane wave''
orthonormal basis $\{\phi_k\}$ for the lattice scalar fields with periodicity
$\phi(x+(2p+1)Le_{\nu})=\phi(x)$ (i.e. the fields in the scalar version
of $\tC_{2p+1}$), given by
\be
\phi_k(x)&=&\frac{1}{\sqrt{\cal N}}\,e^{ik\cdot{}x}
\qquad\qquad{\cal N}=((2p+1)2N)^4 \label{48} \\
k_{\mu}&\in&\frac{2\pi}{a(2p+1)2N}\Big\{-(2p+1)N,-(2p+1)N+1,\dots,
(2p+1)N-1\Big\}
\label{49}
\ee
Note that the volume per $k$ is
\be
\Delta^4k=\Bigl(\,\frac{2\pi}{a(2p+1)2N}\Bigr)^4
=\frac{(2\pi)^4}{a^4{\cal N}}
\label{50}
\ee
Using the plane wave basis, the trace in (\ref{47}) can now be evaluated
as in \cite[v4]{DA(hep-lat)}, leading to
\be
q_{\tU}^{(2p+1)}(x)=q_A(x)\sum_ka^4\Delta^4k\,\hat{I}(ak)+O(a)
\label{51}
\ee
where the summation region for $k$ is (\ref{49}), $\Delta^4k$ is given by
(\ref{50}),
\be
q_A(x)=\frac{1}{32}\,\epsilon_{\mu\nu\rho\sigma}\,\tr(F_{\mu\nu}(x)
F_{\rho\sigma}(x))
\label{52}
\ee
and
\be
\hat{I}(k)=\frac{-3}{8\pi^2}\,\frac{\prod_{\nu}\cos{}K_{\nu}\Bigl(-1+
\sum_{\mu}(1-\cos{}k_{\mu})-\sum_{\mu}\frac{\sin^2k_{\mu}}{\cos{}k_{\mu}}
\Bigr)}{\Big\lb\,\sum_{\mu}\sin^2k_{\mu}+(-1+\sum_{\mu}(1-\cos{}k_{\mu}))^2
\Big\rb^{5/2}}
\label{53}
\ee
and we have exploited the fact that $\tA_{\mu}(x)=A_{\mu}(x)$ for
$x\in\Gamma$.
Changing summation variable from $k$ to $\tilde{k}=ak$ in (\ref{51}) and taking
the $p\to\infty$ limit gives
\be
\lim_{p\to\infty}\,q_{\tU}^{(2p+1)}(x)=q_A(x)\int_{-\pi}^{\pi}d^4\tilde{k}\,
\hat{I}(\tilde{k})+O(a)
\label{54}
\ee
The integral over $\tilde{k}$ in 
this expression was evaluated in \cite{DA(hep-lat)},
and independently in \cite{Suzuki}, and was found to be 1. Recalling
(\ref{30a})--(\ref{30}) we finally get 
\be
\lim_{a\to0}\;q_U(x)=\lim_{a\to0}\,\lim_{p\to\infty}\;q_U^{(2p+1)}(x)
=\lim_{a\to0}\,\lim_{p\to\infty}\;q_{\tU}^{(2p+1)}(x)=q_A(x)
\qquad\quad(x\in\Gamma)
\label{55a} 
\ee
and
\be
\lim_{a\to0}\;Q^f(U)&=&
\lim_{a\to0}\,\lim_{p\to\infty}\;a^4\sum_{x\in\Gamma}q_U^{(2p+1)}(x)
\nonumber \\
&=&\lim_{a\to0}a^4\sum_{x\in\Gamma}(q_A(x)+O(a))
=\int_{T^4}d^4x\,q_A(x)=Q(A)
\label{55}
\ee
This completes the sketch of the proof of the theorem. The latter part of 
the derivation has similarities with, and was inspired by, the calculation 
of the classical continuum limit of the axial anomaly for Wilson--Dirac
fermions done many years ago by W. Kerler \cite{Kerler} and E. Seiler
and I. O. Stamatescu \cite{Seiler}.

The Overlap-Dirac operator determines a lattice Dirac fermion
action $\bar{\psi}D_U\psi$ \cite{Neu1} which has an exact symmetry under
a new kind of lattice chiral transformations \cite{Luscher(PLB)}.
The corresponding axial anomaly (=the infinitesimal chiral jacobian)
was found in \cite{Luscher(PLB)} to be (in the flavour singlet case)
\be
{\cal A}_U(x)=2iq_U(x)
\nonumber
\ee
Its classical continuum limit was calculated in a perturbative setting in
\cite{Kiku} and subsequently in a non-perturbative setting in
\cite[v1]{DA(hep-lat)},\cite{Fuji(NPB)},\cite{Suzuki}. The perturbative
setting was further studied in \cite{Chiu(pert)}.
However, these 
calculations are all problematic in the case where the continuum gauge field
$A_{\mu}(x)$ is topologically non-trivial (e.g. $A_{\mu}(x)$ does not have
a well-defined Fourier expansion in this case.) This case is covered by
the arguments above though: we have found that
\be
q_U(x)=q_A(x)+O(a)
\nonumber
\ee
in complete generality. As we have seen, the locality of the Overlap-Dirac
operator (which was not used in the previous references) is a key ingredient 
in the derivation of this result. 
 
The arguments and results above, which are for SU(N) gauge fields on the
4-torus, can be easily extended to U(1) fields on the 2-torus, thereby 
providing analytic verification of results from numerical studies in
\cite{overlap} where the fermionic topological charge and axial anomaly
were seen to reduce to the 
correct continuum quantities for topologically non-trivial gauge fields
in this setting.

\section{Index of lattice Dirac operators and combinatorial approach to
topological invariants}

In this section a potential mathematical application of the ideas behind
fermionic lattice topological charge to the construction of certain
topological invariants of manifolds is discussed. Consider K\"ahler--Dirac
spinor fields \cite{Kahler,Joos-Rabin} with the ``spacetime'' being an
arbitrary smooth oriented riemannian 4-dimensional manifold $M$. 
In the continuum the fields are
the differential forms on $M\,$, the space of which we denote by
$\Omega(M)=\oplus_{p=0}^4\Omega^p(M)$. (Locally, a p-form
$\omega\in\Omega^p(M)$ is of the form 
$\omega_{\mu_1\dots\mu_p}(x)dx^{\mu_1}\wedge\cdots\wedge{}dx^{\mu_p}$.)
The Dirac operator on $\Omega(M)$ is 
\be
D=d+d^*
\label{61}
\ee
where $d_p:\Omega^p(M)\to\Omega^{p+1}(M)$ is the exterior derivative.
The standard way to define the chirality operator $\Gamma_5$ (analogue of
$\g5$) is $\Gamma_5=(-1)^p$ on $\Omega^p(M)$. This has the chirality 
properties $(\Gamma_5)^2=1$ and $\Gamma_5D=-D\Gamma_5\,$, and it is not 
difficult to show that the corresponding index of $D$ is the Euler 
characteristic $\chi(M)$ of $M$ (see, e.g., \cite[Ch.4]{Nash}).
A discretisation of K\"ahler--Dirac theory
can be obtained via a triangulation $K$ of $M$ \cite{Joos-Rabin}.
In the discrete theory the fields are the cochains of $K$ (=the functions
on the oriented simplexes of $K$), the space of which we denote by
$C(K)=\oplus_{p=0}^4C^p(K)$. The analogue of $d$ is the coboundary operator
$d^K:C^p(K)\to{}C^{p+1}(K)\,$, the Dirac operator is $D^K=d^K+(d^K)^*$
and $\Gamma_5=(-1)^p$ on $C^p(K)$. Using the de Rham theorem it can be shown
that $\index\,D^K=\index\,D=\chi(M)$ so $\index\,D^K$ is a combinatorial
construction of the topological invariant $\chi(M)$. This is nothing new
though; in fact we have just recovered Euler's original combinatorial
construction of $\chi(M)$. Now, there is another way to define $\Gamma_5$
in the continuum, namely by $\Gamma_5=(-1)^s\ast$ where $\ast:\Omega^p(M)
\to\Omega^{4-p}(M)$ is the Hodge star operator and $s=(-1)^{1+p(p-1)/2}$
on $\Omega^p(M)\,$; this also satisfies the chirality properties 
$(\Gamma_5)^2=1$ and $\Gamma_5D=-D\Gamma_5$ and the corresponding index
is again a topological invariant of $M$, namely the Hirzebruch 
{\em signature} $\sigma(M)$ (see, e.g., \cite[Ch.4]{Nash}).
Unlike $\chi(M)$ there is no known combinatorial construction of $\sigma(M)$
-- this is an interesting open problem in mathematics. The problem is
reflected in the fact that there is no natural discretisation of $\ast\,$,
and thereby of $\Gamma_5\,$, in this setting.
However, a natural discretisation is obtained after introducing $\hK\,$: the
cell decomposition of $M$ dual to $K$. We can then consider the doubled
discretisation $\Omega(M)\to{}C(K)\oplus{}C(\hK)$ with
\be
D&\to&D^{K\hK}=\left(\,{D^K \atop 0}\;{0 \atop D^{\hK}}\,\right)
\qquad,\qquad{}
\Gamma_5\to\dg5=\left(\,{0 \atop (-1)^s\ast^K}\;
{(-1)^s\ast^{\hK} \atop 0}\,\right)
\label{62}
\ee
where $\ast_p^K:C^p(K)\to{}C^{4-p}(\hK)$ and 
$\ast_q^{\hK}:C^q(\hK)\to{}C^{4-q}(K)$ are the {\em duality} operators
(see, e.g., \cite{DA(PRL+Rtorsion)}, where topological quantities were
exactly reproduced in a similar discrete setting).
The index of $D^{K\hK}$ can be seen to vanish though; this is essentially 
due to the fact that $\dg5$ is skew-diagonal in (\ref{62}).
This is reminiscent of the vanishing of the index of the usual naive lattice 
Dirac operator. The necessity of introducing $C(\hK)$ 
is reminiscent of the appearance of fermion doubling in the usual
lattice theory. Therefore, in light of the preceding sections, it may be 
possible to obtain a combinatorial construction of $\sigma(M)$ from the
spectral flow of a ``Hermitian Wilson--Dirac operator''
$H_{(m)}^{K\hK}=\dg5(D_{Wilson}^{K\hK}-mf(a))$.\footnote{
Note that in the continuum, $\sigma(M)=\index\,D$
equals minus the spectral flow of the hermitian operator
$H_{(m)}=\Gamma_5(iD-m)$ as $m$ varies from any negative value to any positive 
value.}
The main problem is to find a suitable ``Wilson term'' $W^{K\hK}$
on $C(K)\oplus{}C(\hK)$ for the ``Wilson--Dirac operator''
$D_{Wilson}^{K\hK}=iD^{K\hK}+aW^{K\hK}$ (where $a$ is now the mesh size
of $K$), and a suitable function $f(a)$ in $H_{(m)}^{K\hK}$. (In the usual
lattice setting $f(a)=1/a$.) One idea is to embed $C(K)\oplus{}C(\hK)$
into $C(BK)$ where $BK$=the barycentric subdivision of $K$, and then
try to construct $W^{K\hK}$ from the discrete Laplace operator
$\Delta^{BK}$ on $C(BK)$. 

Finally, we remark that after ``twisting'' by a flat gauge field $A$,
the signature $\sigma_A(M)$ of a 4-manifold $M$ with boundary $N$ is
closely related to the Atiyah--Patodi--Singer rho invariant
$\rho(N,\alpha)$ of a 3-manifold $N$ together with a representation
$\alpha:\pi_1(N)\to{}O(n)$ \cite{APS(II)}. A combinatorial construction of 
the signature invariant $\sigma_A(M)$ would in all likelihood lead to a
combinatorial construction of the rho invariant as well.

\vspace{1ex}
\noindent {\em Acknowledgements.} I thank
Ting-Wai Chiu and everyone else involved in the running and organisation
of Chiral'99 for a very enjoyable and stimulating conference, and the 
National Center for Theoretical Sciences in Hsinchu for hospitality
and financial support in the weeks leading up to Chiral'99.
Also, I thank Martin L\"uscher for discussions
on the classical continuum limit, from which I benefited
greatly, and for hospitality during a visit to DESY.
Thanks also go to Varghese Mathai for discussions on the signature invariant.
The support of an ARC postdoctoral fellowship is gratefully acknowledged.


\begin{thebibliography}{XXX}


\bibitem{AS}
M. Atiyah and I. Singer, Ann. Math. {\bf 87} (1968) 484

\bibitem{L}
M. L\"uscher, Comm. Math. Phys. {\bf 85} (1982) 39

\bibitem{PhillipsU(1)}
A. Phillips, Ann. Phys. {\bf 161} (1985) 399

\bibitem{PS}
A. Phillips and D. Stone, Comm. Math. Phys. {\bf 103} (1986) 599

\bibitem{U(1)problem}
G. 't Hooft, Phys. Rev. Lett. {\bf 37} (1976) 8; Phys. Rev. D {\bf 14}
(1976) 3432; R. Jackiw and C. Rebbi, Phys. Rev. Lett. {\bf 37} (1976) 172;
C. Callan, R. Dashen and D. Gross, Phys. Lett. B {\bf 63} (1976) 334;
Phys. Rev. D {\bf 17} (1978) 2717

\bibitem{Wilson}
K. Wilson, Phys. Rev. D {\bf 14} (1974) 2445

\bibitem{Smit}
J. Smit and J. Vink, Nucl. Phys. B {\bf 286} (1987) 485

\bibitem{Itoh}
S. Itoh, Y. Iwasaki and T. Yoshi\'e, Phys. Rev. D {\bf 36} (1987) 527

\bibitem{DA(herm)}
D. Adams, hep-lat/9907005

\bibitem{local}
P. Hern\'andez, K. Jansen and M. L\"uscher, Nucl. Phys. B 
{\bf 552}, 363 (1999)

\bibitem{Neu(bound)}
H. Neuberger, hep-lat/9911004

\bibitem{overlap}
H. Neuberger and R. Narayanan, Phys. Lett. B {\bf 302} (1993) 62;
Nucl. Phys. B {\bf 412} (1994) 574; {\bf 443} (1995) 305

\bibitem{Neu1}
H. Neuberger, Phys. Lett. B {\bf 417} (1998) 141

\bibitem{Neu2}
H. Neuberger, Phys. Lett. B {\bf 427} (1998) 353

\bibitem{GW}
P. Ginsparg and K. G. Wilson, Phys. Rev. D {\bf 25} (1982) 2649

\bibitem{laliena}
P. Hasenfratz, V. Laliena and F. Niedermayer, Phys. Lett. B {\bf 427}
(1998) 125

\bibitem{Luscher(PLB)}
M. L\"uscher, Phys. Lett. B {\bf 428} (1998) 342

\bibitem{Fuji(NPB)}
K. Fujikawa, Nucl. Phys. B {\bf 546} (1999) 480

\bibitem{Edwards}
R. Edwards, U. Heller and R. Narayanan, Nucl. Phys. B {\bf 522} (1998) 285 

\bibitem{Seiler(Top)}
F. Karsch, E. Seiler and I. Stamatescu, Nucl. Phys. B {\bf 271} (1986) 349

\bibitem{Vranas}
R. Narayanan and P. Vranas, Nucl. Phys. B {\bf 506} (1997) 373

\bibitem{Gattringer}
C. Gattringer, I. Hip and C. Lang, Nucl. Phys. B {\bf 63} (1998) 498

\bibitem{Hernandez}
P. Hern\'andez, Nucl. Phys. B {\bf 536} (1998) 345

\bibitem{Chiu(PRD)(PRD)}
T.-W. Chiu, Phys. Rev. D {\bf 58} (1998) 074511; {\bf 60} (1999) 114510

\bibitem{Woit}
P. Woit, Phys. Rev. Lett {\bf 51} (1983) 638

\bibitem{Luscher(private)}
M. L\"uscher, private communication

\bibitem{prep}
D. Adams, in preparation

\bibitem{DA(hep-lat)}
D. Adams, hep-lat/9812003

\bibitem{Suzuki}
H. Suzuki, Prog. Theor. Phys. {\bf 102} (1999) 141

\bibitem{Kerler}
W. Kerler, Phys. Rev. D {\bf 23} (1981) 2384

\bibitem{Seiler}
E. Seiler and I. Stamatescu, Phys. Rev. D {\bf 25} (1982) 2177;
{\bf 26} (1982) 534 (E)

\bibitem{Kiku}
Y. Kikukawa and A. Yamada, Phys. Lett. B {\bf 448} (1999) 265

\bibitem{Chiu(pert)}
T.-W. Chiu and T.-H. Hsieh, hep-lat/9901011

\bibitem{Kahler}
E. K\"ahler, Rendiconti di Matematica {\bf 21} (1962) 425;
W. Graf, Ann. Inst. Henri Poincar\'e A {\bf 29} (1978) 85

\bibitem{Joos-Rabin}
P. Becher and H. Joos, Z. Phys. C {\bf 15} (1982) 343;
J. Rabin, Nucl. Phys. B {\bf 201} (1982) 315

\bibitem{Nash}
C. Nash, {\it Differential topology and quantum field theory.}
(Academic Press, London, 1976)

\bibitem{DA(PRL+Rtorsion)}
D. Adams, Phys. Rev. Lett. {\bf 78} (1997) 4155; hep-th/9612009

\bibitem{APS(II)}
M. Atiyah, V. Patodi and I. Singer, Math. Proc. Camb. Phil.
Soc. {\bf 78} (1975) 405


\end{thebibliography}
\end{document}